\documentclass[a4paper,12pt]{article}
\pdfoutput=1 

\usepackage[T1]{fontenc} 

\usepackage{draft}
\usepackage{putex} 
\usepackage{soul}

\usepackage{cite}

\usepackage[
      colorlinks=true,
      linkcolor=black,
      urlcolor=blue,
      filecolor=black,
      citecolor=red,
      ]{hyperref}
      
\usepackage{graphicx}
\usepackage{xcolor}
\usepackage{amsfonts,amsmath,amssymb,bm}
\usepackage{verbatim}
\usepackage{array}
\usepackage{mathrsfs}
\usepackage{mathtools}
\usepackage{yfonts}
\usepackage{dsfont}
\usepackage{bbm}
\usepackage{colonequals}
\usepackage{amscd}
\usepackage{slashed}

\usepackage{float}
\usepackage{afterpage}

\usepackage{subcaption}

\usepackage{wrapfig}

\usepackage{suffix,mathtools,cancel,bbm}

\usepackage{multirow}

\hypersetup{
    colorlinks,%
    citecolor=blue,%
    filecolor=blue,%
    linkcolor=blue,%
    urlcolor=blue
}

\usepackage{tikz}
\usetikzlibrary{positioning}

\renewcommand{\title}[1]{\vbox{\center\LARGE{#1}}\vspace{5mm}}
\renewcommand{\author}[1]{\vbox{\center#1}\vspace{5mm}}

\makeatletter
\def\leftrightarrowsfill@{\arrowfill@\leftrarrows\Rrelbar\lrightarrows}
\newcommand{\xleftrightarrows}[2][]{\ext@arrow 3399\leftrightarrowsfill@{#1}{#2}}
\makeatother

\begin{document}

\institution{HA}{Department of Physics, Harvard University, 17 Oxford Street, Cambridge MA 02138, USA}

\institution{BHI}{Black Hole Initiative, Harvard University, Cambridge MA 02138, USA}

\institution{MI}{Dipartimento di Fisica, Universita' di Milano, via Celoria 16, 20133 Milano MI, Italy}

\institution{INFN}{INFN, Sezione di Milano, Via Celoria 16, I-20133 Milano, Italy
}

\title{The spectrum of near-BPS Kerr-Newman black holes and the ABJM mass gap}

\authors{Matthew Heydeman\worksat{\HA,\BHI} and Chiara Toldo\worksat{\HA, \MI, \INFN} \footnote[0]{mheydeman@fas.harvard.edu, chiaratoldo@fas.harvard.edu}
}

\abstract{Supersymmetric rotating 1/16-BPS black holes in $AdS_4 \times S^7$ are expected to capture the average degeneracy of BPS states in the dual ABJM superconformal theory for given fixed charges. This has been successfully demonstrated for the superconformal index using complexified black hole metrics, but a naive Gibbons-Hawking calculation of the actual degeneracies in the low temperature limit is invalid due to large quantum fluctuations of the near horizon AdS$_2$ metric. We argue that in a particular mixed grand/canonical ensemble, these fluctuations of the near-BPS Kerr-Newman black holes are described by a version of the $\mathcal{N}=2$ super-Schwarzian theory with $SU(1,1|1)$ symmetry. Using this description as well as properties of ABJM, we recover the large $N$ superconformal index and find a characteristic  ``mass gap'' of order $N^{-3/2}$ between the 1/16-BPS states and the lightest near BPS state. We further make a prediction for the operator dimension spectrum above the gap in the large $N$, low $T$ limit. Our results are consistent with the Bekenstein-Hawking formula at large energies, random matrix theory at low energies, and the microscopic index.

}

\date{}

\maketitle

{
  \hypersetup{linkcolor=black}
  \tableofcontents
}

\pagebreak

\setcounter{page}{1}

\section{Introduction}

A basic fundamental hypothesis about quantum gravity is that the Bekenstein-Hawking entropy may be regarded thermodynamic or coarse-grained measure of the number of states in the microscopic black hole Hilbert space. While there has historically been issues in making this idea precise, the landmark work of \cite{Strominger:1996sh} showed that string theory can give a microscopic interpretation of black hole entropy in terms of counting strings on D-branes. This approach leverages both dualities and supersymmetry to compute not the degeneracies of generic black holes, but instead what could be thought of as the supersymmetric index of the (charged) black hole. Such indices essentially only receives contributions from ground states, counting bosonic or fermionic states with opposite sign. This applies to extremal (supersymmetric) black holes, having large charges and area but vanishing Hawking temperature: the asymptotic growth of the index indeed reproduces the Hawking area formula exactly, implying an exponentially large ground state degeneracy. Away from this supersymmetric limit, no exact counting is known, but understanding the bulk gravitational physics at low temperatures (in the so-called ``near-extremal'' regime) has become an important paradigm for the study of quantum black holes.

The extremal limit of a black hole has some curious features. One is that the Third Law of Thermodynamics is violated due to the exponentially large ground state degeneracy. This could be explained if the ground states were protected by supersymmetry as in the Strominger-Vafa example, but this is not particularly generic. Further, even if we only consider supersymmetric or BPS black holes, the index is not guaranteed to be equal to the degeneracy due to Bose-Fermi cancellation. 

A more serious potential problem in the near-extremal limit was pointed out more than thirty years ago in \cite{Preskill:1991tb}, where it was argued that the semiclassical thermodynamic description of a black hole breaks down close to extremality, because the heat capacity at low enough temperatures becomes small (while the area remains macroscopic). Since standard thermodynamics relates the heat capacity with fluctuations in temperature and entropy, this means that temperature fluctuations become larger than the black hole temperature itself. In other words, at very low temperature, the emission of even a single Hawking quantum can drastically alter the temperature of the near-extremal black hole, violating the assumption of thermodynamic equilibrium. In this regime, adding quantum corrections is necessary to determine the true low temperature nature of the black hole spectrum. This conundrum has two potential resolutions: either there is an exact degeneracy at $T=0$, followed by a gap in the spectrum of states (below which semiclassical thermodynamics does not apply), or the exact energy and entropy are smoothly reduced below the Bekenstein-Hawking result due to quantum fluctuations, removing the large ground state degeneracy.

To address the issue of quantum effects near extremality, the work of \cite{Ghosh:2019rcj,Iliesiu:2020qvm,Heydeman:2020hhw} employed the gravitational path integral in a low temperature expansion. While the full higher dimensional path integral may not be done directly (and may not even be well defined), it is possible to identify a set of modes of the metric which become strongly coupled at low temperatures and may not be treated using only the semiclassical approximation. The set of modes and their fluctuation determinants depend sensitively on which symmetries are preserved by the exact extremal solution, including supersymmetry and other global symmetries. This information may be repackaged into a two-dimensional theory consisting of JT gravity\cite{Teitelboim:1983ux,Jackiw:1984je} coupled to matter, or a supersymmetric extension thereof. The advantage of passing to this two dimensional description is that the leading contribution to the path integral comes from geometries with the topology of the hyperbolic disc (with exact $SL(2,\mathbb{R})$ symmetry), while the strongly coupled modes of the metric are essentially the boundary fluctuations of this disc (boundary reparametrizations which are not gauge equivalent to $SL(2,\mathbb{R})$ transformations) \cite{Maldacena:2016upp}. The low temperature effective field action of the boundary mode is Schwarzian derivative of the boundary profile, and the path integral may actually be evaluated exactly\cite{Stanford:2017thb,Mertens:2017mtv}, including for isometry groups more general than $SL(2,\mathbb{R})$. The same result may be found without explicitly using a two-dimensional truncation by instead regulating (by means of a near-extremal solution) certain zero modes present in the path integral in the near-horizon region \cite{Iliesiu:2022onk,Banerjee:2023quv}.

The results of the above analysis reveal that the semiclassical thermodynamics is indeed invalid at sufficiently low temperatures. Without supersymmetry, the large ground state degeneracy is removed by low temperature fluctuations and the density of states goes smoothly to zero as the energy above extremality is decreased. However, in a charge sector which permits supersymmetric black holes, the spectrum contains $\exp(S_{BH})$ exactly degenerate supersymmetric ground states separated from the non-BPS states by a mass gap, where the value of the gap depends on the parameters of the semiclassical solution. This spectrum is consistent with both quantum statistical mechanics for non-BPS black holes, but also the results of the microscopic counting when such a comparison can be made.

The gravitational path integral may thus be performed in certain situations, and it provides coarse grained information about the spectrum of black hole microstates (subject to corrections that are subleading in the extremal entropy and the temperature). While the JT/Schwarzian description does not rely on supersymmetry, the results for supergravity are interesting because one may make direct contact with microscopic constructions of black holes and AdS/CFT found in string theory. An example of this is the near-BPS analysis of AdS$_5$ black holes \cite{Boruch:2022tno}, corresponding to $1/16$ states of SYM. The relevant 2d theory is $\mathcal{N}=2$ JT gravity, which describes the finite temperature breaking of the $SU(1,1|1)$ superisometry found in the near-horizon region. With appropriate input from the full Type IIB supergravity, this theory reproduces the extremal entropy and index of $\mathcal{N}=4$ SYM at large $N$, but also predicts a gap of order $1/N^2$ in the spectrum. The non-BPS states above this gap are unprotected, so finding this spectrum is a problem in strongly coupled four-dimensional supersymmetric gauge theory\footnote{There is some hope that features of the gravity answer, such as the gap\cite{Chang:2023zqk}, near extremal free energy \cite{Cabo-Bizet:2024gny}, and random matrix statistics\cite{Chen:2024oqv} may already be seen at weak gauge coupling.}.

In the present paper we extend the above ideas to the case of AdS$_4$ supergravity, giving us a chance to compare the behaviour with the both the flat space and AdS$_5$ results. For the 4d asymptotically flat Kerr solution, it was shown that the ground state degeneracy is lifted by quantum corrections \cite{Kapec:2023ruw,Rakic:2023vhv}\footnote{See also \cite{Ghosh:2019rcj} \cite{Maulik:2024dwq} \cite{Kapec:2024zdj,Kolanowski:2024zrq} for related work on quantum corrections to rotating black holes in different dimensions.}. In contrast with flat space, AdS$_4$ Kerr-Newman black holes may preserve supersymmetry, which means we may hope to find a discrete set of BPS states in agreement with the superconformal index of the dual CFT$_3$. Furthermore, a strongly coupled gravity calculation in the near-BPS regime allows us to make predictions about the (average) spectrum of operator dimensions in the CFT above the BPS bound.

In more detail, we work with theories of 4d $\mathcal{N} = 2$ minimal gauged supergravity, which is the ``universal'' consistent truncation of 11d supergravity on homogeneous 7d Sasaki-Einstein manifolds ($SE_7$), and admit AdS$_4$ supersymmetric black hole solutions, whose entropy can be reproduced via a microscopic computation in the dual theory (see \cite{Zaffaroni:2019dhb} for a review). In AdS$_4$ there are two classes of such black holes, discovered and studied first in \cite{Romans:1991nq,Kostelecky:1995ei,Caldarelli:1998hg}, both preserving one quarter of supersymmetry (which is 1/16-th for the case of $S^7$). One branch of solutions, the supersymmetric Kerr-Newman AdS$_4$ black hole, is characterized by the presence of electric charge and a nonvanishing angular momentum (the static limit being singular). When lifted up to 11d on $S^7$, these solutions have an ABJM \cite{Aharony:2008ug} field theory dual, and their entropy can reproduced by the computation of the superconformal index (SCI), see for instance \cite{Kim:2009wb,Choi:2019zpz,Nian:2019pxj,BenettiGenolini:2023rkq}. The other branch instead preserves supersymmetry via the so-called partial topological twist \cite{Witten:1988xj}, and has a smooth static limit. The entropy of these black holes was reproduced via the computation of the twisted index of ABJM, starting with the work of \cite{Benini:2015eyy}. The twisted index (as well as an analogous twisted partition function) is really counting states of the SCFT when quantized on a spatial Riemann surface; it is a different observable than the CFT partition function in radial quantization. Since our goal in this work is to focus on spinning conformal primaries, we will discuss new phenomenon for the twisted case in a sequel \cite{Heydeman:2024fgk}. 

\noindent Schematically, our proposal for the Kerr-Newman ABJM black hole is:
\begin{figure}[H]
	\begin{center}
	\includegraphics[scale=0.45]{"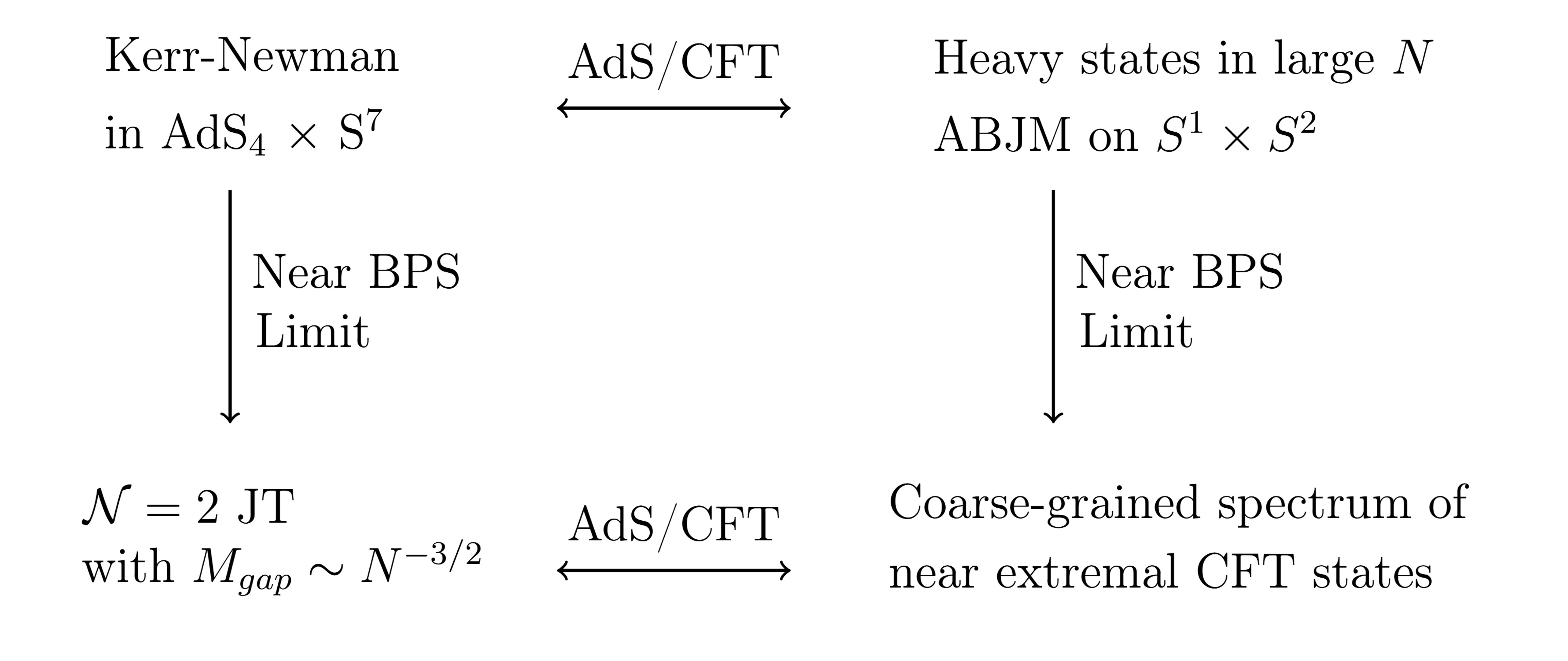"}
 \end{center}
	\end{figure}
\vspace{-1 cm}

In this work, we will argue that the low temperature expansion of the gravitational path integral for the AdS$_4$ Kerr-Newman black hole reduces to a version of the $\mathcal{N}=2$ Schwarzian theory in the near BPS regime\footnote{Work on near-extremal perturbations of BPS and non-BPS black holes in theories of gauged and ungauged supergravity appeared in \cite{Castro:2021wzn}. The authors pointed out a possible instability in the scalar spectrum for a certain class of (nonminimally coupled) dyonic black holes, which we do not treat in this paper.}. In principle, one must either perform a dimensional reduction of the gravitational theory in the near horizon region to AdS$_2$ and isolate the low-energy modes that contribute to the one-loop determinant via temperature dependent terms, or identify these same modes by expanding the full nonextremal metric at low temperature. While this can be made completely explicit in simpler examples\cite{Heydeman:2020hhw}\footnote{Non-supersymmetric reductions to JT may be found in \cite{Almheiri:2014cka,Nayak:2018qej,Moitra:2018jqs,Moitra:2019bub,Iliesiu:2020qvm}
.}, both of these options are technically unwieldy for 11-dimensional spacetimes. Instead, we will use semiclassical thermodynamics and knowledge of the symmetries of the extremal solution to argue for the appropriate ``infrared'' effective field theory, understanding that non-universal features of the higher dimensional theory have been integrated out\cite{Iliesiu:2020qvm}. Our proposal (based on the properties of the 11d classical solution) for the $\mathcal{N}=2$ Schwarzian can be seen as the boundary term of a 2d $\mathcal{N}=2$ JT supergravity in the near horizon region, and the parameters of this theory are fixed by the higher dimensional black hole and serve as EFT coefficients. It is important that the JT gravity we consider exists in the context of a bigger model; the spectrum depends on several discrete and continuous parameters that must be computed, and the output is only the approximate (though fully quantum mechanical) large $N$ answer.

In order for the near-BPS expansion to be valid, we must consider black holes in a particular ensemble in which the near AdS$_2$ solution provides the dominant saddle point contribution as we scale the inverse temperature $\beta$. To isolate the extremal black hole and its small fluctuations (and not say, thermal AdS or some horizonless geometry with small charges), we work in a mixed ensemble as done in \cite{Boruch:2022tno}; in this ensemble one linear combination of the black hole charges is kept fixed and large, while charge fluctuations of the other linear combination are treated grand canonically with a chemical potential $\alpha$. This potential $\alpha$ is chosen so that it couples to the infrared R-charge preserved by the black hole, and is interpreted as the boundary holonomy of a gauge field in JT supergravity corresponding to local $U(1)_R \in SU(1,1|1)$ symmetry. In the near BPS ensemble, the general form of the field theory partition function contains a sum of saddles, a feature which also appears in the AdS$_2$ gravity theory in a precise way as a sum over gauge bundles of AdS$_2$. This is reflected in the sum over the parameter $m$ in the formula for the Super-Schwarzian partition function \cite{Stanford:2017thb}, which is 1-loop exact and takes the form:
\begin{equation}\label{very_first_formula}
    Z_{\mathcal{N}=2 \textrm{ JT}}(\beta, \alpha ; r, \vartheta) = \sum_{m \in \frac{1}{r}\cdot \mathbb{Z}} e^{i r \vartheta m} \left( \frac{2 \cos(\pi (\alpha + m))}{\pi (1-4(\alpha+m)^2)} \right) e^{S_0 + \frac{2\pi^2}{\beta M_{SU(1,1|1)}}(1-4(\alpha +m)^2)} \, ,
\end{equation}
where we isolated in round brackets the contribution of the 1-loop determinant, in contrast to the exponential terms which have a purely semiclassical origin as the on-shell free energy in the Gibbons-Hawking approach. In particular, $S_0$ is the extremal entropy and $M_{SU(1,1|1)}$ is an energy scale where the semiclassical thermodynamics breaks down, as in \cite{Preskill:1991tb}. While these parameters may be computed from the classical solution, the $SU(1,1|1)$ Schwarzian theory becomes strongly coupled at $\beta \sim M_{SU(1,1|1)}^{-1}$ and the 1-loop effects cannot be ignored.

The effective 2d theory is furthermore characterized by two free discrete parameters $r, \vartheta$ that are not singled out from classical low energy (AdS$_2$) considerations. The abelian $U(1)_R$ symmetry allows for fractional charge quantization which is the case for $r \neq 1$ and $r^{-1} \in \mathbb{Z}$. Further, 2d abelian gauge theory admits a possible topological $\vartheta$-angle term with coefficient we call $\vartheta$. As can be seen from the partition function, both these parameters dictate affect how we sum over gauge bundles and so are associated to the spectrum of charges. These parameters must be computed in the bigger gravity theory and play an important role in determining the behaviour of the density of states, as we will see in the main body of the paper. While the value of the $\vartheta$ parameter can be determined by the 4d supergravity model under consideration (which depend on the 7-dimensional $SE_7$ used in the 11d truncation\cite{Genolini:2021qbi}), for the other parameter we use a minimal input from the dual CFT.

While much of what we say applies more generally to M-theory backgrounds which permit AdS$_4 \times SE_7$ vacua described by 4d $\mathcal{N}=2$ supergravity (and have 3d superconformal field theory duals arising arising from coincident M2-branes probing a cone over $SE_7$), we focus on the well known $S^7$ compactification which is dual to the $SU(N)$ $k=1$ ABJM which describes $N$ coincident M2 branes. When this theory is compactified on $S^1_{\beta} \times S^2$ with background potentials for rotations and R-symmetry, the path integral has an interpretation as a grand canonical partition function which counts the operator content of this theory. While this partition function is difficult to evaluate at strong coupling, a specialization of chemical potentials leads to the superconformal index of ABJM, and this quantity is computable at large $N$ and matches the entropy of supersymmetric black holes with spin and R-charge\cite{BenettiGenolini:2023ucp}. Our work from the bulk black hole point of view is consistent with this large $N$ index, but the solvability of the Schwarzian theory allows us to make predictions for the behavior of the spectrum of unprotected operators with $\Delta > \Delta_{BPS}$.

\begin{figure}[h]
	\begin{center}
	\includegraphics[scale=0.43]{"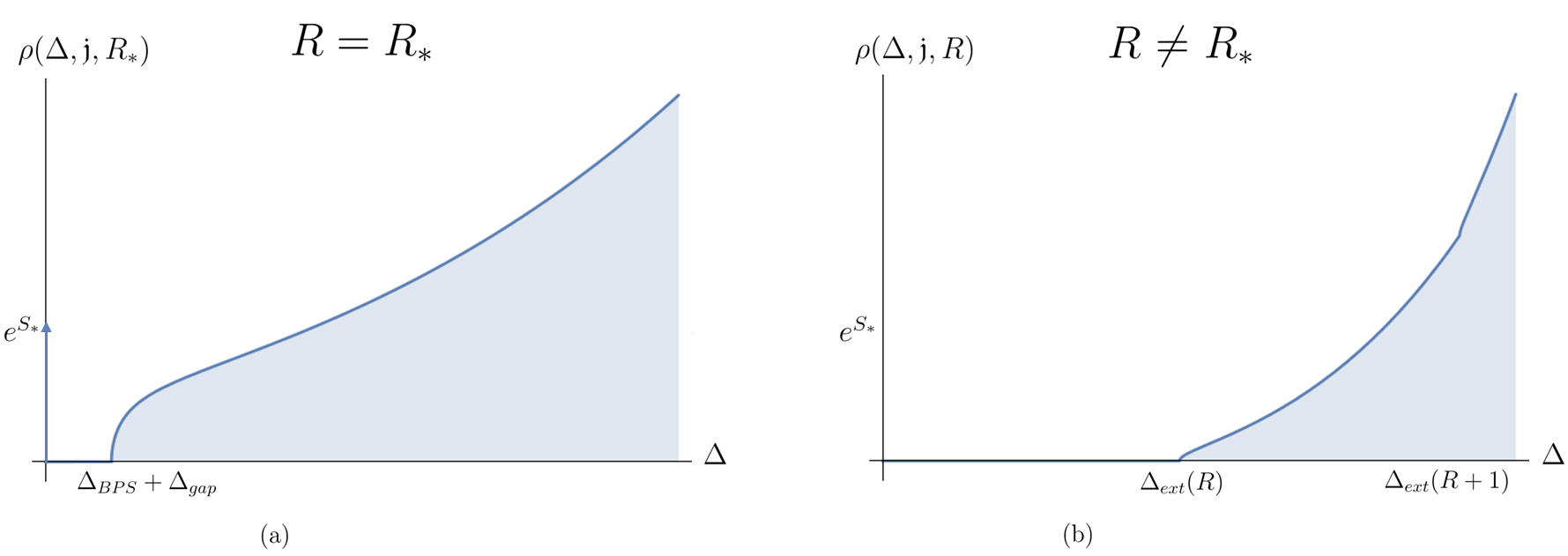"}
 \end{center}
	\caption{Graph (a) displays the density of states, Eq.~\eqref{eq:BPSDOS}, for a near BPS black hole ($R=R_*$), which displays a Dirac delta function associated to the extremal BPS solution, and a mass gap after which a continuous density of states starts. Graph (b) is the density of states for the same theory, Eq.~\eqref{eq:nonBPSDOS}, but now for nonsupersymmetric black holes with $R \neq R_*$, for which the density of states goes to zero as $T \rightarrow 0$; the theory predicts a vanishing number of extremal black hole microstates with these charges. Even though the solutions are not BPS, all states must organize into representations of $\mathcal{N}=2$ supersymmetry; the kink in the blue line represent superpartners with the same charges which enter at their respective gap energy. \label{Fig0}}
	\end{figure}

As our main result, we use the higher dimensional solutions to compute the parameters of \eqref{very_first_formula} and show that the quantum corrected density of states, extracted by performing the Laplace transform, contains both a discrete BPS degeneracy and a continuous spectrum of long multiplets, interpreted as a coarse-grained average of spinning black hole microstates. For ABJM, the 1/16th-BPS and non-BPS states are separated by a mass gap of order $N^{-3/2}$, so the quantum corrected gravity result predicts no (exponential number of) microstates with operator dimensions in this region. We derive the near-BPS spectrum of states above the gap, providing a prediction for the spectrum of near-BPS states in the dual theory at strong coupling. This behavior is displayed in in Fig \ref{Fig0}.

The paper is organized as follows. We start with the description of the electrically charged Kerr-Newman AdS$_4$ black hole, discussing their thermodynamics, supersymmetry properties, and near horizon geometry (section \ref{sec:KN}). After briefly explaining the main strategy to compute the quantum corrected density of states, in section \ref{sec3:FT} we elucidate the mapping of the ABJM parameters to the effective near horizon JT-theory. The computation of the corrected density of states entails the calculation of the low temperature expansion of the on-shell action for black holes, which is performed in \ref{expansion_freeenergy}. After doing this, all ingredients are put together in section \ref{sec:quantumKN4}, where the quantum-corrected answer for black hole spectrum is given.

\section{Kerr-Newman AdS$_4$ black holes\label{sec:KN}}

We give a brief summary of the black hole solutions under consideration, focusing on $\mathcal{N} =2$ gauged supergravity in four dimensions. The bosonic part of its action is just the Cosmological Einstein-Maxwell theory 
\begin{equation}
\label{eq:4dN2sugra}
    S = \frac{1}{16 \pi G} \int d^4x \sqrt{g} \left(R + 6 - F^2 \right) \,,
\end{equation}
where $R$ is the Ricci scalar and $F$ is the field strength of a $U(1)$ gauge field (graviphoton). We adopt the conventions of \cite{BenettiGenolini:2023rkq} and set the cosmological constant to $\Lambda = -3/l_{AdS}^2 = -3$. $\mathcal{N}=2$ gauged supergravity  \cite{Freedman:1976aw,Fradkin:1976xz,Romans:1991nq} is itself a consistent truncation of 11d supergravity on $S^7$ \cite{deWit:1986oxb}. In these conventions we have the following relation between the 4d Newton's constant $G$ and the rank $N$ of the gauge group in the field theory dual
\begin{equation}
    G = \frac{3}{2\sqrt{2}}N^{-\frac32} \, .
\end{equation}
Because the black hole solutions we consider do not excite additional Kaluza-Klein scalars or other matter from the M-theory perspective, a version of our results would be valid for compactifications of M-theory on a more general Sasaki-Einstein 7-fold \cite{Gauntlett:2007ma}, dual to 3d superconformal field theories with $\mathcal{N}=2$ supersymmetry. One generalization is the orbifold reduction of M-theory on $S^7/\mathbb{Z}_k$, where $k$ is the level of the dual ABJM Chern-Simons matter theory; here we take $k=1$ and it plays no further role in the gravity analysis.

The theory contains additional fermions, namely a doublet of gravitini (which here are packaged into a complex one). A semiclassical analysis in gravity involves setting all the fermions to zero in the classical solution. Because we are focused on near BPS black holes (which include the exactly BPS ones), we will need to explain the supersymmetry of a solution, which is signified by the existence of a supercovariantly constant Killing spinor. This spinor is annihilated by the local supersymmetry variation of the gravitino, and the Killing spinor equation is:
\begin{align} \label{Killing-spinor}
    \left (\nabla_\mu - i A_\mu +\frac12 \Gamma_\mu +\frac{i}{4}F_{\nu \rho}\Gamma^{\nu \rho}\Gamma_\mu \right ) \epsilon = 0 \,,
\end{align}
where
\begin{equation}
\nabla_{\mu} \epsilon = \partial_{\mu} \epsilon - \frac14 w_{\mu}^{ab} \gamma_{ab} \epsilon \,.
\end{equation}
While a purely classical analysis of the entropy in gravity cannot distinguish theories with additional fermions and theories without, such as the cosmological Einstein-Maxwell theory \eqref{eq:4dN2sugra}, or the $\mathcal{N} =2$ supergravity theory which comes with the additional gravitini, the presence of the extra fields does change the 1-loop thermodynamics. As explained in \cite{Heydeman:2020hhw,Boruch:2022tno}, the fermion determinants in the near-extremal limit guarantee the supersymmetry of the resulting spectrum and are required to reproduce the superconformal index and other interesting features such as the mass gap\footnote{In studying quantum corrections more generally, one might complain that this logic should apply to all fluctuating fields of the 11d metric, and not just the components that appear in our truncation. One can argue \cite{Iliesiu:2020qvm,Iliesiu:2022onk}  that near extremality, these additional fluctuations contribute to the $\log(S)$ corrections to the entropy. Since we want to focus on the $\log(T)$ corrections only, we will work at leading order in the extremal entropy $S_0$ with the understanding that additional corrections may be included in a systematic way, see for instance \cite{Sen:2012kpz} and related work. }.

Through the consistent truncation argument, the 11-dimensional black holes we wish to study appear as charged, rotating solutions of the effective four-dimensional theory presented above. The relevant rotating Kerr-Newman AdS solutions with electric charge are known from \cite{Carter:1968ks} (their supersymmetry properties were subsequently studied in \cite{Kostelecky:1995ei,Caldarelli:1998hg}) and they take the following form:
\begin{eqnarray} \label{Kerr_new_gen}
    ds^2 & = & -\frac{\Delta_r \Delta_{\theta}}{V(r, \theta) \, \Xi^2} dt^2 + U(r,\theta) \left(\frac{dr^2}{\Delta_r}+ \frac{d\theta^2}{\Delta_{\theta}} \right)+ \nonumber \\
    & + & \sin^2 \theta \left( V(r, \theta) d\phi + a \Delta_{\theta}   \frac{\Delta_r -(1+r^2) (r^2+a^2)}{U(r,\theta) \, \Xi^2} dt\right)^2 \, ,
\end{eqnarray}
with vector field
\begin{equation}
    A = \frac{m r \sinh \delta}{U(r,\theta) \, \Xi} (\Delta_{\theta} dt - a \sin^2 \theta d \phi) + \gamma dt \, , 
\end{equation}
and the definitions,
\begin{equation}
    \Delta_r = (r^2 +a^2) (1+r^2) -2m r \cosh \delta + m^2 \sinh^2 \delta \,, \nonumber
\end{equation}
\begin{equation}
    \Delta_{\theta} = 1-a^2 \cos^2 \theta \,,  \qquad \Xi = 1-a^2  \,,\qquad U (r, \theta) = r^2 +a^2 \cos^2 \theta \,, \nonumber 
\end{equation}
\begin{equation}
     V (r, \theta) = \frac{\Delta_{\theta} (r^2+a^2)^2 - a^2 \sin^2 \theta \Delta_r}{U(r, \theta) \, \Xi^2} \,.
\end{equation}
The range of coordinates is $ 0 \leq \theta < \pi$, $r,t>0$ and $0 \leq \phi < 2 \pi$. The solution is parameterized by the quantities $m,a,\delta $, which are related to the mass, angular momentum and charge of the black hole, as shown in the next section. The parameter $\gamma$ is a constant that will be fixed later on demanding regularity of the Wick-rotated solution. In what follows we elaborate on the thermodynamic quantities and supersymmetry properties, following somewhat closely \cite{BenettiGenolini:2023rkq}. 

\subsection{Thermodynamic quantities and supersymmetric limit}
The solution \eqref{Kerr_new_gen} represents a spinning black hole with temperature
\begin{equation}
    T = \frac{\Delta_r' (r_+)}{4 \pi(r_+^2 +a^2)} \,,
\end{equation}
while the angular velocity $\Omega$ at the horizon is
\begin{equation}
    \Omega = \frac{a (1+ r_+^2)}{r_+^2+ a^2}\,.
\end{equation}
The horizon surface $r=r_+$ is a Killing horizon of the field
\begin{equation}
V = \frac{\partial}{\partial t} + \Omega \frac{\partial }{\partial \phi}\,.
\end{equation}
Then, the electrostatic potential $\Phi_e$ is computed as
\begin{equation}
    \Phi_e = \imath_V A_{r=r_+}- \imath_V A_{r\rightarrow \infty}   = \frac{m \sinh \delta r_+}{ r_+^2 +a^2}\,.
\end{equation}
Regularity of the gauge field in the Wick rotated solution $t \rightarrow i \tau$ imposes that $\imath_V \mathcal{A}_{r=r_+}=0$, fixing $\gamma = -\Phi_e$. Finally, we report the expression for the entropy of the black hole, computed from the area of the event horizon
\begin{equation}
    S =  \frac{\pi}{G} \frac{r_+^2 +a^2}{1-a^2}\,,
\end{equation}
and the conserved quantities, computed via the standard methods of holographic renormalization \cite{Papadimitriou:2005ii}:
\begin{equation}
\label{eq:KerrCharges}
  M =\frac{m \cosh \delta }{G \, \Xi^2}\,, \qquad J = a M = \frac{a m \cosh \delta }{G \, \Xi^2}\,, \qquad Q = \frac{m \sinh \delta}{G \, \Xi} \,. 
\end{equation}
The renormalized on-shell action, computed from
\begin{equation}
    I = I_0 + \frac{1}{8 \pi G} \int_{\partial M} d^3x \left( K -2 -\frac12 R_3 \right) \, ,
\end{equation}
is
\begin{equation}
\label{eq:onshellaction}
    I = \frac{\beta}{2 G \Xi} \left(m \cosh \delta - \frac{r_+ m^2 \sinh^2 \delta}{r_+^2 +a^2} - r_+ (r_+^2 +a^2)  \right) \, ,
\end{equation}
and it obeys the quantum statistical relation
\begin{equation}
    I = - S + \beta (M - \Omega J - \Phi_e Q)\,.
\end{equation}

The solution described above admits a BPS limit. The supersymmetry condition obtained from \eqref{Killing-spinor} implies the relation $ M = J +Q $ which in turn yields the following constraint between the parameters $a$ and $\delta$
\begin{equation} \label{susy_cond_int}
   a = \coth \delta -1\,.
\end{equation}
Imposing, on top of this, that $\Delta_r$ has a real double root (extremality condition), we get the following constraint on the horizon radius $r_+$ and mass parameter $m$: 
\begin{equation}
r_+^2 = \coth \delta -1 \, , \qquad m^2 = \frac{\cosh^2 \delta} {e^{\delta} \sinh^5 \delta} \,.
\end{equation}
Therefore the extremal BPS  solution is labelled by one free parameter $\delta$. From now on we will denote with the symbol $*$ all the extremal BPS quantities, which are then
\begin{equation}\label{BPSvalues}
r_*^2 = a_* = \coth \delta -1 \, , \qquad m_* = \frac{\cosh \delta}{e^{\delta/2} \sinh^{5/2} \delta} \,.
\end{equation}

\subsection{Near horizon geometry and symmetries}

The near horizon geometry for the BPS solution found above is obtained by performing the following diffeomorphism, in which the parameter $\lambda$ is taken to zero 
\be
r\rightarrow r_++\lambda R_0 R,\;\;\;t\rightarrow\frac{\bar{T} R_0}{\lambda},\;\;\;\phi \rightarrow \bar{\Phi}+\Omega_0\frac{\bar{T} R_0}{\lambda},
\label{eq:NHkerr}
\ee
with
\begin{equation}
   \Omega_0=\frac{\Xi a }{(a^2+r_0^2)}\,, \qquad  R_0^2 = \frac{r_+^2 +a^2}{\Delta_0} \,,
    \label{eq:mextaext}
\end{equation}
where the parameter $\Delta_0$ will be defined momentarily.

The effect of the shift in the angle in \eqref{eq:NHkerr} is to reach the frame co-moving with the horizon. If we plug (\ref{eq:NHkerr}) in (\ref{Kerr_new_gen}), after taking the decoupling limit  $\left(\lambda\rightarrow 0\right)$ with $R$, $T$, $\theta$ and $\bar{\Phi}$ fixed, we find \cite{Compere:2012jk}
\begin{equation}\label{NHEKAdS}
	\begin{split}
ds^2=\Gamma(\theta)\Big[-R^2d\bar{T}^2+\frac{dR^2}{R^2}+\bar{\alpha}(\theta)^2 d\theta^2\Big]+\bar{\gamma}(\theta)(d\bar{\Phi}+K Rd\bar{T})^2,
	\end{split}	
\end{equation}
and
\begin{equation}\label{gaugeAdS}
A = f(\theta) (d\bar{\Phi} + K R d\bar{T}) - \frac{e}{K} d\bar{\Phi}\,,
\end{equation}
where 
\begin{equation}
\Gamma(\theta)=\frac{U(r_+,\theta)}{\Delta_0}\, ,\quad
	\bar{\alpha}(\theta)=\frac{\Delta_0^{1/2}}{\Delta_{\theta}^{1/2}} \, ,\qquad K=\frac{2ar_+\Xi}{\Delta_0 (r_+^2+a^2)}\, ,
 \end{equation}
 and
 \begin{equation}
	\bar{\gamma}(\theta)=\frac{\Delta_{\theta}(r_0^2+a^2)^2\sin^2\theta}{U(r_+,\theta)\Xi^2},
	\label{eq:gammathetaalpha}
\end{equation}
and finally
 \begin{equation}
	e= \frac{Q}{\Delta_0} \frac{r_+^2-a^2}{r_+^2+a^2}\,, \qquad f = \frac{(r_+^2 + a^2) Q (r_+^2-a^2 \cos^2 \theta)}{2 U(r_+,\theta) \Xi a r_+}\,.
\end{equation}
We defined
 \begin{equation}
\Delta_{\theta} = 1-a^2 \cos^2 \theta\,, \qquad U(r_+,\theta) = r_+^2 +a^2 \cos^2 \theta\,, \qquad \Delta_0 = 1 + a^2 +6r_+^2\,.
\end{equation}
In the previous formulae, $r_+$ is just $r$ evaluated at the extremal BPS horizon $r_*$ in \eqref{BPSvalues}.

The near-horizon geometry is a fibered product of AdS$_2$ and a two-sphere. Notice that the well-known Near Horizon Extremal Kerr (``NHEK'') geometry \cite{Bardeen:1999px,Guica:2008mu} is found by setting $Q=0$, and by reinstating the AdS radius $l$ and taking it to infinity.

The configuration \eqref{NHEKAdS}-\eqref{gaugeAdS} enjoys the following isometries
\be \nonumber
\zeta_0 = R \partial_R - \bar{T} \partial_{\bar{T}} \, , \qquad \zeta_{-1} = \partial_{\bar{T}} \, , \qquad \zeta_1 = \frac12 \left( \frac{1}{R^2} +\bar{T}^2 \right)\partial_{\bar{T}} - (\bar{T} R) \partial_R + \frac{1}{R} \partial_{\bar\Phi} \, ,
\ee
\be
\qquad L_0 = \partial_{\bar{\Phi}} \, .
\ee
The vectors $\zeta_{-1}, \zeta_0, \zeta_1$ obey the $SL(2,R)$ commutation relations, therefore the near horizon geometry exhibits $SL(2,R) \times U(1)$ symmetry. The Schwarzian $U(1)$ is a combination of this last rotation combined with a gauge transformation of $A$, and due to the presence of supersymmetry\footnote{The susy constraint \eqref{susy_cond_int} in the previous section comes from the integrability condition, studied in \cite{Caldarelli:1998hg}, which provide a necessary condition to preserve susy.}, the $SL(2, R) \times U(1)$ symmetry then completes to the superalgebra $SU(1,1|1)$, as shown in detail for instance in \cite{Hristov:2013spa,Benini:2015eyy}.  The heuristic logic for this is that the superconformal index of the dual field theory counts BPS representations of the supersymmetry algebra generated by $\mathcal{Q}$ and $\mathcal{Q}^\dagger$\cite{BenettiGenolini:2023rkq},
\begin{equation}
    \{ \mathcal{Q}, \mathcal{Q}^\dagger \} = H - J - \frac12(R_1 + R_2 + R_3 +R_4) \, .
\end{equation}
By viewing $H$ as the Hamiltonian in radial quantization and thus the dilatation generator, and $R\equiv J + \frac12(R_1 + R_2 + R_3 +R_4)$, these supercharges and bosonic generators combine with the remaining generators of $SL(2,\mathbb{R})$ to form a $SU(1,1|1)$ algebra. We will later explain the relationship between these charges and those present in the black hole solution which is a truncation of the 11d problem.

If we turn on finite temperature, this symmetry is naively broken at the level of the classical metric. However, at sufficiently low temperatures, the bosonic and fermionic fluctuations around the AdS$_2$ geometry are controlled by the existence of this symmetry\cite{Maldacena:2016upp}, and the presence of a gravitational soft mode whose effective action is (a supersymmetric extension of) the Schwarzian derivative plays a crucial role in the gravity analysis to follow.

\section{Mixed Ensemble and the Low Temperature Expansion }
\label{sec3:FT}
As we explain below, the Schwarzian corrections to the semiclassical gravity partition function are most readily computed in a ``mostly canonical'' or mixed ensemble, where almost all charges are fixed, and a Legendre transform is performed for the remaining linear combination, resulting in a partition function which depends on the inverse temperature $\beta$ and a single chemical potential, which we denote with $\alpha$. In this ensemble, the large fixed charges imply the classical near-extremal black hole is the dominant saddle point. The Schwarzian corrections are then the most dominant $\beta$ and $\alpha$ dependent 1-loop determinants around this background. We omit many of the details of how these 1-loop determinants are computed (instead focusing on the details specific to ABJM), but the strategy is similar to that in \cite{Iliesiu:2020qvm,Heydeman:2020hhw,Boruch:2022tno} using JT gravity. Of course, the same kinds of results may be obtained without any specific reference to JT \cite{Iliesiu:2022onk,Banerjee:2023quv, Kapec:2023ruw,Rakic:2023vhv,Kolanowski:2024zrq}, but we do not attempt this here.

Actually, in the large $\beta$ limit with fixed $\alpha$, one must actually include a sum of saddle points corresponding to different ``winding configurations'' for the bulk fields which couple to $\alpha$. This requirement may already be seen in field theory computations of the index, as argued in \cite{Aharony:2021zkr,BenettiGenolini:2023rkq}, and the same is true for the near-BPS partition function (though we cannot compute this directly in field theory). To see why we must include a sum, one notes that the abstract form of the partition function as a trace over a Hilbert space naturally incorporates a $2\pi n$ periodicity under shifts of the chemical potentials. If we approximate this trace as a contribution from a single saddle point (as is usually done in gravity), this periodicity is lost. This in turn translates into a violation of charge quantization, but it can be restored by summing over saddles. The precise nature of this sum depends on the underlying properties of the SCFT and the chosen ensemble.

Therefore, in what follows we first abstractly manipulate the 3d SCFT partition function such that it is amenable to a semiclassical expansion. We will show (without computing it directly), that this expansion may be mapped to the partition function of the related super-Schwarzian theory based on the symmetries determined in the previous section. Assuming then that the dominant contribution to the black hole partition function comes from modes in the throat region (as was demonstrated concretely in the above referenced examples), we use the known result for the super Schwarzian 1-loop corrections to determine the leading behavior of the partition function at sufficiently low temperatures. This will allow us to extract the low energy spectrum. A simple check of this proposal is to compute the superconformal index ($\alpha = \frac12$ in this language) directly via the Schwarzian theory, and we show this is exactly the large $N$ limit of \cite{BenettiGenolini:2023rkq}.

\subsection{The field theory partition function \label{subsec4.1}}
The conformal field theory we consider is the $U(N)_k \times U(N)_{-k}$ ABJM theory for $k=1$ which is strongly coupled. Due to non-perturbative effects from monopole operators, the 3d $\mathcal{N}=6$ superconformal symmetry manifested in the Lagrangian is enhanced to $\mathcal{N}=8$ superconformal symmetry. This leaves the superconformal algebra $\mathfrak{osp}(8|4)$ which has the bosonic subalgebra $\mathfrak{so}(3,2) \times \mathfrak{so}(8)$. The vacuum state of this theory preserves all symmetries and in thus dual to M-theory on AdS$_4 \times$ S$^7$. 

We can abstractly consider the thermal partition function in the grand canonical ensemble (GCE) of $\mathcal{N}=8$ ABJM by placing the theory on $S^1 \times S^2$ in the presence of background fields\cite{Kim:2009wb,Choi:2019zpz,Nian:2019pxj,BenettiGenolini:2023rkq}. Equivalently, in the Hamiltonian formalism we may radially quantize the theory on $S^2$ and take the trace refined by background chemical potentials. States in the Hilbert space $\mathcal{H}_{S^2}$ are labeled by their eigenvalues under the maximal Cartan subalgebra of the bosonic symmetry algebra, which is $U(1)_H \times U(1)_J \times U(1)^4_R$ for the Hamiltonian $H$, angular momentum $J$, and the 4 $R$-charges $R_a$, $a = 1,\dots 4$. In our conventions, we take the spectrum of $J$ and $R_a$ to be half integral, and $J = R_a$ \textrm{mod} $1$ for all states. We have chosen our conventions to match \cite{BenettiGenolini:2023rkq}, which analyzed the phases of the ABJM superconformal index and the matching to bulk complex euclidean black holes.

Our goal to is to organize the field theory partition function in a form which can be compared with the extremal limit of a bulk calculation. In the context of microstate counting in AdS/CFT, this procedure is usually applied to the superconformal index. However, since we are not computing an index, but instead a thermal partition function, we can consider the most general GCE partition function:
\begin{equation}
\label{eq:GCEZ}
    Z(\beta, \Omega, \Phi_a) = \textrm{Tr}_{\mathcal{H}_{S^2}} e^{- \beta H + \beta \Omega J + \sum_{a=1}^4 \beta \Phi_a R_a} \, ,
\end{equation}
which depends on the inverse temperature and 5 conjugate potentials $(\beta, \Omega, \Phi_a)$ for the angular velocity and electric potentials. At strong coupling, this quantity generically cannot be computed in field theory, but we may attempt a semiclassical approximation in the dual gravity description.

While we are ultimately interested in non-BPS states, specializing the partition function to an index acts as a guide for how we should define the near-BPS partition function. This specialization would involve placing one complex constraint on the 5 potentials, and the resulting (temperature independent) partition function can be chosen to count states annihilated by a supercharge and its conjugate $\mathcal{Q}$, $\mathcal{Q}^\dagger$ such that
\begin{equation}
\label{eq:susycommutator}
    \{ \mathcal{Q}, \mathcal{Q}^\dagger \} = H - J - \frac{1}{2}(R_1 + R_2 + R_3 + R_4) \, .
\end{equation}
The BPS states we count in the index are thus annihilated by the combination of bosonic generators on the right hand side. From the point of view of the gravity dual, it is possible to motivate why extremal ($\beta \rightarrow \infty$) black holes may satisfy this condition. We can regard $H$ as the dilation generator inside an $SL(2,\mathbb{R})$ subalgebra, and $J + \frac12 \sum_a R_a$ as a $U(1)_R$ symmetry which does not commute with supersymmetry. The full superconformal algebra we preserve in the BPS sector is $\mathfrak{su}(1,1|1)$, and this is the near horizon symmetry of BPS AdS$_4$ rotating black holes with electric charges \cite{Hristov:2011ye,Hristov:2013spa}. While we expect these black holes to give the dominant contribution to the index (and more general partition functions) at strong coupling, there are other bulk geometries such as thermal AdS which still preserve the $\mathfrak{su}(1,1|1)$. To make contact with our proposed bulk calculation which focuses on fluctuations of near-BPS black holes, we must change our ensemble (in both field theory and gravity) to focus on states that contribute in this limit, essentially states with charges of order $G^{-1}$, as in \eqref{eq:KerrCharges}.

Let us manipulate the general GCE partition function to obtain a quantity which only receives contributions from states with large charges close to the BPS bound. Starting with \eqref{eq:GCEZ}, we may insert the supersymmetry relation to obtain
\begin{equation}
\label{eq:GCEQQb}
    Z(\beta, \Omega, \Phi_a) = \textrm{Tr}_{\mathcal{H}_{S^2}} e^{- \beta \{\mathcal{Q},\mathcal{Q}^\dagger \} + \beta (\Omega-1) J + \sum_{a=1}^4 \beta (\Phi_a-\frac12) R_a} \, .
\end{equation}
We now introduce new charges which commute with supersymmetry,
\begin{equation}
\label{eq:flavorcharges}
    \mathfrak{j} = J + \frac14 \sum_{a=1}^4 R_a \, , \qquad \mathfrak{q}_{1,2,3} = R_{1,2,3} - R_4 \, ,
\end{equation}
as well as new reduced imaginary chemical potentials which absorb the factors of $\beta$ and track the deviation of the potentials from their extremal values
\begin{equation}
    \tau = \frac{\beta}{2 \pi i}(\Omega - 1) \, , \qquad  \lambda_{1,2,3} + \frac{\tau}{4} = \frac{\beta}{2 \pi i}(\Phi_{1,2,3} - \frac12) \, .
\end{equation}
Finally, we will eliminate $\Phi_4$ in terms of a new chemical potential $\varphi$ and its reduced version $\alpha$:
\begin{equation}
\label{eq:BPSpotential}
    \varphi = \Phi_1 + \Phi_2 + \Phi_3 + \Phi_4 -\Omega - 1 \, , \qquad \quad \alpha = \frac{\beta}{4\pi i}\varphi \, .
\end{equation}

The potential $\alpha$ is useful in defining the specialization which leads to the index:
\begin{equation}
    \textrm{Susy:} \quad \beta \left(\sum_a \Phi_a - \Omega - 1 \right ) = 2 \pi i n_0 \, , \quad n_0 \in 2 \mathbb{Z} + 1 \quad \Longrightarrow \quad \alpha = \frac{n_0}{2} \, .
\end{equation}
We will see shortly why this special choice of $\alpha$ leads to the usual superconformal index. A more general value of $\alpha$ captures the approach to the BPS surface as we send $\beta \rightarrow \infty$ \cite{Larsen:2020lhg,BenettiGenolini:2023rkq}. In terms of the dual black hole description, $\alpha$ is the chemical potential corresponding to a boundary condition for a gauge field in the AdS$_2$ region \cite{Heydeman:2020hhw,Boruch:2022tno}.  

Using these definitions, the full GCE partition function \eqref{eq:GCEQQb} may be written as:
\begin{align}
    Z(\beta, \tau, \lambda_{1,2,3}, \alpha) = \textrm{Tr}_{\mathcal{H}_{S^2}} e^{- \beta \{\mathcal{Q},\mathcal{Q}^\dagger \} + 2\pi i \tau \mathfrak{j}  +  \sum_{i=1}^3 2\pi i \lambda_i \mathfrak{q}_i +  2 \pi i \alpha (2 R_4)} \, . \label{eq:ZGCE2}
\end{align}
When thinking about this expression as the abstract partition function of the boundary ABJM theory, we can see how this reduces to the usual superconformal index in the conventions of \cite{BenettiGenolini:2023rkq}. In field theory,
\begin{align}
    J = R_a \, \, \textrm{ (mod 1)},\, \, \,   \textrm{for all $a=1,\dots 4$} \,.
\end{align}
This means that when $n_0$ is an odd integer so that $\alpha = \frac{n_0}{2}$ is half integral, we have
\begin{align} \label{spin_stat_CFT}
    e^{2\pi i \alpha (2 R_4)} = e^{2\pi i n_0 J} = (-1)^{2J} \, .
\end{align}
This factor plays the role of the $(-1)^F$ for this refined index due to spin-statistics, and the thermal partition function
\begin{equation}
    Z(\beta, \tau, \lambda_{1,2,3}, \alpha = \frac{n_0}{2}) \equiv \mathcal{I}(\tau; \lambda_{1,2,3}) \, , 
\end{equation}
is precisely the refined index of \cite{BenettiGenolini:2023rkq}.

\subsection{Charge quantization and periodicity of $Z$}
In our original variables, the relationship $J = R_a$ \textrm{mod} $1$ implies a particular periodicity of the partition function under imaginary shifts of the chemical potentials,
\begin{align}
    Z(\beta, \Omega, \Phi_a) &= \textrm{Tr}_{\mathcal{H}_{S^2}} e^{- \beta H + \beta (\Omega + \frac{2 \pi i}{\beta}n_\Omega) J + \sum_{a=1}^4 \beta (\Phi_a + \frac{2 \pi i}{\beta}n_a) R_a} \, , \\ 
    &= \textrm{Tr}_{\mathcal{H}_{S^2}} e^{- \beta H + \beta \Omega  J + \sum_{a=1}^4 \beta \Phi_a  R_a + 2\pi i (n_\Omega J + \sum n_a R_a)} \, .
\end{align}
Therefore, the partition function is invariant under arbitrary (constrained) shifts satisfying 
\begin{equation}
\label{eq:integershifts}
    n_\Omega + \sum_a n_a = 2 \mathbb{Z} \, .
\end{equation}
This kind of invariance is automatic in the field theory definition of the partition function via the trace and follows from charge quantization. However, in the dual supergravity description in which we approximate this function by the exponentiated action of a classical bulk solution, the periodicity is lost and charge quantization is seemingly no longer enforced \cite{Aharony:2021zkr}.  One proposal to fix this issue is to improve the approximation of $Z$ by including, in addition to the dominant saddle, a sum over a family of bulk solutions which satisfy the same boundary conditions but differ in the choice of bundle for the bulk gauge fields corresponding to $\Omega, \Phi_a$. As bulk solutions, these complex black holes are smooth and have a subleading action compared to the original black hole, but their effect becomes important in the near-extremal limit \cite{Heydeman:2020hhw,Boruch:2022tno}.

In addition to simplifying the partition function to \eqref{eq:ZGCE2}, our change of variables also decouples the constraint on the integer shifts \eqref{eq:integershifts}. From \eqref{eq:flavorcharges}, we see that for all states, $\mathfrak{j},\mathfrak{q}_i$ have an integral spectrum. While $R_4$ is still half integral, the periodicity is doubled in \eqref{eq:ZGCE2}, so that we have the following shift invariance:
\begin{equation}
\label{eq:periodicity}
    Z(\beta, \tau +4 n_\Omega, \lambda_{1,2,3}+ n_{1,2,3}, \alpha+m) = Z(\beta, \tau, \lambda_{1,2,3}, \alpha) \, ,
\end{equation}
for unconstrained $n_\Omega, n_i, m \in \mathbb{Z}$.

In addition to the periodicities described above, there is another subtle quantum effect which is sensitive to the discreteness of the charge spectrum but not manifest in the gravity description. Thinking of $J + \frac12 \sum_a R_a$  as an infrared $U(1)_R$ charge which appears in \eqref{eq:susycommutator}, the R-charge spectrum of BPS states may depend on the charges of elementary fields relative to those of the preserved supercharges $\mathcal{Q}$, $\mathcal{Q}^\dagger$. In particular, it is possible for the index to vanish in theories with an infrared $SU(1,1|1)$ symmetry if the theories contain fractionally charged ground states. This is known to happen in simple holographic theories such as the $\mathcal{N}=2$ SYK model \cite{Fu:2016vas,Heydeman:2022lse}  as well as $\mathcal{N}=2$ JT gravity \cite{Stanford:2017thb} which is relevant for this work.

Using the table from \cite{BenettiGenolini:2023rkq} which lists the weights of elementary fields under the Cartan chosen above, it is possible to determine if this phenomenon of fractional charges is present. The infrared R-charge of $\mathcal{Q}$ is $\frac12$, and one can see that no elementary field has an R-charge which is a fraction of this. Of course, it is already known that the superconformal index of ABJM does not vanish, so one might not have been worried about this possibility. Nevertheless, we know of no general argument that guarantees the index does not suffer large cancellations with this small amount of susy preserved, though there are arguments available in other examples\cite{Sen:2009vz,Mandal:2010cj,Benini:2015eyy}. Even if one considers a holographic theory with large supersymmetric black holes in the gravity dual, it could be possible to have a large number of black hole microstates which exactly cancel in  the index (as happens in $\mathcal{N}=2$ SYK).

\subsection{The sum over saddles and the mixed ensemble \label{subs:sumOversaddles}}
So far, we have rewritten the GCE partition function in a convenient form, \eqref{eq:ZGCE2} and determined the allowed periodicity \eqref{eq:periodicity}. In principle, one could proceed with a bulk calculation in which we approximate the partition function as a sum over bulk saddles which asymptote to euclidean AdS$_4 \times$ S$^7$ and have the correct holonomies and periodicities. However, in the application to near-BPS black holes, this strategy has a a few disadvantages. 

The black hole solutions we consider arise as truncations of the 4d theory obtained by compactifying eleven-dimensional supergravity on S$^7$. These truncations do not have the full $SO(8)$ gauge symmetry or scalar sector, but it is easier to find a class of explicit solutions with fewer $U(1)^4$ charges. However, the field theory quantity we are trying to compute does not enforce this convenient restriction on the charges, so generically a sum over saddles will run outside of the class of geometries we are trying to consider. Since we want to focus on the low temperature fluctuations around a specific black hole with large charges tuned to extremality, we will transform the partition function into one in the \emph{mixed ensemble} (ME) in which most of the global charges are fixed except for those conjugate to the temperature and BPS chemical potential \eqref{eq:BPSpotential} \cite{Boruch:2022tno}.

We begin by approximating the GCE partition function as a sum over shifted saddles bulk, as motivated in the previous section,
\begin{align}
    Z(\beta, \tau, \lambda_{1,2,3}, \alpha) &= \textrm{Tr}_{\mathcal{H}_{S^2}} e^{- \beta \{\mathcal{Q},\mathcal{Q}^\dagger \} + 2\pi i \tau \mathfrak{j}  +  \sum 2\pi i \lambda_i \mathfrak{q}_i +  2 \pi i \alpha (2 R_4)}  \\
    &\approx \sum_{n_\Omega, n_1, n_2,n_3,m} Z_{\textrm{one-loop}}e^{-I_{GCE}(\beta, \tau + 4 n_\Omega, \lambda_1 + n_1, \lambda_2 + n_2,\lambda_3 + n_3, \alpha +m)} \,.
\end{align}
Here, $I_{GCE}$ is the (properly regularized) bulk gravitational action evaluated on a solution with the given boundary conditions for the metric and gauge fields. In practice, these solutions will be black hole metrics with the dominant action given these boundary conditions, but in principle these are not the only bulk objects which contribute. $Z_{\textrm{one-loop}}$ are the various one loop determinants of matter and Kaluza-Klein fields around a given saddle. In this regard, much attention in the literature has been given to the log(Area) corrections, see for instance \cite{Sen:2012kpz,Sen:2012cj}. Motivated by studies of finite temperature AdS$_2$ holography via JT gravity and its supersymmetric counterparts, more recent work has demonstrated the existence of nontrivial log($T$) corrections which occur when a higher dimensional black hole approaches extremality \cite{Iliesiu:2022onk,Banerjee:2023quv,Kapec:2023ruw,Rakic:2023vhv,Banerjee:2023gll,Maulik:2024dwq,Karan:2024gwf}. In this section, we will ignore the 1-loop determinants, which will be determined by this kind of gravity reasoning later.

To make contact with this bulk calculation, we will work in an ensemble with the following fixed charges defined in \eqref{eq:flavorcharges}:
\begin{equation}
     R_1=R_2=R_3=R_4 \quad \rightarrow \quad \mathfrak{q}_1 = \mathfrak{q}_2 = \mathfrak{q}_3 = 0\, , \quad \textrm{and}\quad  \mathfrak{j} = J + R_4 \, \, \, \textrm{fixed}\, .
\end{equation}
We Laplace transform with respect to the periodic potentials to produce a trace in the Hilbert space of the charges fixed above, 
\begin{align}
    Z(\beta, \mathfrak{j}, \mathfrak{q}_i = 0, \alpha) &= \textrm{Tr}_{(\mathfrak{j}\, ; \, \mathfrak{q}_i = 0)} \, e^{- \beta \{\mathcal{Q},\mathcal{Q}^\dagger \} + 2 \pi i (2\alpha R_4)} \, \\
    &= \int_0^1 d\tau \int_0^1 d\lambda_1 \int_0^1 d\lambda_2 \int_0^1 d\lambda_3 \, e^{-2\pi i \tau \mathfrak{j} }  Z(\beta, \tau, \lambda_{1,2,3}, \alpha) \,.
\end{align}
In semiclassical gravity, it is standard to do this change of ensemble by including the appropriate extra boundary terms \cite{Brown:1992bq,Hawking:1995ap}. Once we have fixed the charge of a dynamical $U(1)$ field in the bulk, we no longer may sum over any shifted solutions because these correspond to imaginary shifts of the holonomy. Thus the bulk quantity we want to compute in the mixed ensemble is
\begin{align}
\label{eq:mixedZ}
    Z(\beta, \mathfrak{j}, \alpha) = \sum_{m \in \mathbb{Z}} Z_{\textrm{one-loop}}e^{-I_{GCE}(\beta, \mathfrak{j}, \alpha+m)-2\pi i \tau \mathfrak{j}} \equiv \sum_{m \in \mathbb{Z}} Z_{\textrm{one-loop}}e^{-I_{ME}(\beta, \mathfrak{j}, \alpha+m)} \,.
\end{align}
As we have emphasized, we have arrived at this expression by manipulating the abstract field theory partition function. It is not protected by supersymmetry for $\alpha$ away from the supersymmetric value, and we do not have techniques for computing the relevant mixed ensemble saddles or their fluctuation determinants. However, for large $\beta$ we expect this quantity to be computable in the bulk via the description in terms of JT gravity. The existence of supercharges \eqref{eq:susycommutator} which imply an $\mathfrak{su}(1,1|1)$ symmetry means that we should compare our field theory ansatz with exact partition function of $\mathcal{N}=2$ JT gravity \cite{Stanford:2017thb,Mertens:2017mtv}, which we reproduce here for convenience:
\begin{equation}\label{FE_JT_with_theta}
    Z_{\mathcal{N}=2 \textrm{ JT}}(\beta, \alpha ; r, \vartheta) = \sum_{m \in \frac{1}{r}\cdot \mathbb{Z}} e^{i r \vartheta m} \frac{2 \cos(\pi (\alpha + m))}{\pi (1-4(\alpha+m)^2)}e^{S_0 + \frac{2\pi^2}{\beta M_{SU(1,1|1)}}(1-4(\alpha +m)^2)} \, .
\end{equation}
The properties and spectrum of this theory are reviewed in \cite{Boruch:2022tno}. There are two discrete parameters $r$ and $\vartheta$. The parameter $r$ may be $1$ or fractional and is related to whether the microscopic theory contains fractionally charged states; equivalently it is the radius of the compact boson describing $U(1)_R$ gauge fluctuations at the boundary of the AdS$_2$ disc. We have already determined $r=1$ from the microscopic partition function. 

The parameter $\vartheta$ arises from a potential topological term for the $U(1)_R$ gauge field. It is related to a mixed anomaly between the R-symmetry and the fermion number \cite{Kapec:2019ecr}. Charge conjugation invariance requires $\vartheta=0$ or $\vartheta = \pi$, and the later case indicates an anomaly (analogous to a 1d version of the parity anomaly). This anomaly term does not appear in the semiclassical black hole analysis, and if it were present, we would have exponentially fewer BPS states than those of the large $N$ index in \cite{BenettiGenolini:2023rkq}. In four-dimensions, a topological term of the form $S(\vartheta) \sim \vartheta \int F\wedge F$ would have the correct properties when reduced around the extremal solution. In the particular case of ABJM (or more generally any Sasaki-Einstein 7-fold compactification of M-theory), such a 4d $\vartheta$ term cannot arise \cite{Genolini:2021qbi}. Even if it did, because the black holes we consider only have electric charges, but not magnetic, this term evaluates to zero on our solution. We will fully analyze the consequences of such a term in a companion paper 
 using a different class of solutions \cite{Heydeman:2024fgk}.

\subsection{Low-$T$ expansion of the action and $\mathcal{N}=2$ JT \label{expansion_freeenergy}}

We work in an ensemble of fixed $\mathfrak{j}$ and fixed $\mathfrak{q} =0$ and we expand the on-shell action around the 1/4 BPS solution with $T=0$ and $\varphi =0$, turning on small nonzero values of $\varphi, T \ll 1 $ and keeping $\alpha = \frac{\beta \varphi}{4 \pi i}$ fixed\footnote{The parameter $\alpha$ is the chemical potential that remains finite in the JT limit. In other words, it is the holonomy of the $U(1)$ JT gauge field. Setting $\alpha =1/2$ is equivalent to imposing the susy constraint.}. 

Taking the ensemble of fixed $\mathfrak{j}, \mathfrak{q}$ (the  mixed ensemble) involves adding to the classical on-shell action a boundary term\cite{Brown:1992bq,Hawking:1995ap}, such that 
\begin{equation}
I_{ME} (\beta, \mathfrak{j}, \alpha) = I(\mathfrak{q} =0, \mathfrak{j}, \alpha) + 2 \pi i \, \tau \, \mathfrak{j} \,,
\end{equation}
where the grand-canonical action is evaluated at values of chemical potentials such that it is dominated by $\mathfrak{q} =0$ and $\mathfrak{j}$ fixed.

We expand the on-shell action \eqref{eq:onshellaction} for a generic solution around the BPS values in eq. \eqref{BPSvalues}, defining the deviations from the BPS black hole as
\begin{equation}
(r_*, a_*, q_*) \rightarrow (r_* + \delta r, a_* + \delta a, q_* + \delta q) \, .
\end{equation}
The parameters $\delta r, \, \delta q , \, \delta a$ are then expressed in terms of the temperature $T$ and the quantity $\varphi$ (and $\mathfrak{j}$) in a general power series:
\begin{equation}
    \delta r = \delta r_{\varphi} \varphi + \delta r_T T + \delta r_{\varphi^2} \varphi^2 + \delta r_T^2 T^2 +\delta r_{\varphi,T} \varphi T +...
\end{equation}
with similar expressions for $a$ and $q$. For our purposes working with the terms of second order in $\varphi, T$ will be sufficient. Even so, the nonlinear relationship between the horizon radius, the mass, and the charge lead to a somewhat complicated dependence of the coefficients on the underlying parameters. More explicitly,
\begin{equation}
    \delta r = \frac{ (c_3 a_2-a_3 c_2) T + (a_3 b_2 - a_2 b_3) \varphi}{X} +...
\end{equation}
\begin{equation}
    \delta a =  \frac{ a_3 c_1 T -a_3 b_1 \varphi}{X}  + ...
\qquad 
    \delta q =  \frac{ a_2 c_1 T - a_2 b_1 \varphi}{X} + ...
\end{equation}
where $X = a_2 b_3 c_1 - a_2 b_1 c_3 + a_3 b_1 c_2 - a_3 b_2 c_1  $ and

\begin{equation}
  a_2 = \frac{4 e^{-\frac{\delta_* }{2}} \left(e^{\delta_* /2} \left(e^{4 \delta_* }+7\right) \sqrt{\sinh \delta_* }+\left(-4 e^{2 \delta_* }+e^{4 \delta_* }+3\right) \sqrt{\coth \delta_* -1}\right)}{\left(e^{2 \delta_* }-3\right)^3 \left(e^{2 \delta_* }+1\right)  \sqrt{\sinh \delta_* } (\coth \delta_* -1)^{3/2}} \nonumber \, ,
\end{equation}
\begin{equation}
a_3 = \frac{\left(e^{2 \delta_*}-1\right)^2}{2 \left(e^{2 \delta_*}-3\right)^2 } \, , \qquad b_1 = \frac{e^{\delta_* } \left(10 e^{2 \delta_* }+e^{4 \delta_* }-7\right) \sinh \delta_*  \sqrt{\coth \delta_* -1}}{8 \pi  \left(e^{2 \delta_* }+1\right)} \nonumber \, , 
\end{equation}
\begin{equation}
    \quad 
    b_2 = -\frac{e^{-\frac{5 \delta_*}{2}} \left(e^{2 \delta_*}-3\right) \sqrt{e^{2 \delta_*}-1}}{2 \sqrt{2} \pi  \left(e^{2 \delta_*}+1\right) \sinh ^{\frac{5}{2}}(\delta_*) (\coth (\delta_*)-1)^{5/2}} \, , \nonumber \qquad
    b_3 = \frac{e^{-\frac{\delta_*}{2}} (\sinh \delta_*-\cosh \delta_*)}{2 \pi  \sinh ^{\frac{3}{2}}(\delta_*) (\coth \delta_*-1)^{5/2}} \, ,  
\end{equation}
\begin{equation}
c_1= \frac{\left(e^{2 \delta_* }-3\right) \sqrt{\coth \delta_* -1}+2 e^{\delta_* /2} \left(\cosh \delta_* \sqrt{\text{csch}\delta_* }-2 \sqrt{\sinh \delta_* }\right)}{2 \left(e^{2 \delta_* }+1\right) (\coth \delta_* -1)^{3/2}} \, , \nonumber
\end{equation}
\begin{equation}
c_2= - \frac{\left(\left(e^{2 \delta_* }-3\right) \sqrt{\coth \delta_* -1}\right)+8 e^{\delta_* /2} \left(\cosh \delta_* \sqrt{\text{csch}\delta_* }-\sqrt{\sinh \delta_* }\right)}{2 \left(e^{2 \delta_* }+1\right) (\coth \delta_* -1)^{3/2}} \, ,  \nonumber
\end{equation}
\begin{equation}
c_3= \frac{\tanh \delta_*}{\sqrt{\coth \delta_*-1}} \, , 
\end{equation}

Here we reported only the first order coefficients in the expansion.
The expressions for the second order coefficients are highly involved and we do not report them here, but can be easily computed with software for symbolic manipulation such as Mathematica. We can then insert these values in the on-shell action and expand it. Keeping only terms up to second order in $(\varphi, T)$ we get the result
\begin{equation} \label{FE_exp_mix1}
    I_{ME} (\beta, \mathfrak{j}, \varphi) =  -S_* - \beta \frac{\varphi}{2} Q_* - \frac{2 \pi^2}{ \beta M_{gap}} \left( 1+ 4 \left(\frac{\beta \varphi}{4\pi} \right)^2 \right)\,,
\end{equation}
where the starred quantities assume these values:
\begin{equation}
    S_* = \frac{\pi}{G} \frac{r_*^2+a_*^2}{1-a_*^2} = \frac{2 \pi }{\left(e^{2 \delta_*}-3\right) G}\,, \,\, \qquad Q_* = \frac{\sinh ^2\delta_*}{G \sqrt{e^{\delta_* } \sinh ^5\delta_*} (2-\coth\delta_*)}  \,,
\end{equation}
and 
\begin{eqnarray} \label{MGAP}
    M_{gap}^{-1} & = & \frac{a^{3/2}}{G (1-a_*) (1+6a_* +a_*^2 )} \nonumber \\
    & = & \frac{(\coth \delta_* -1)^{3/2}}{G (2-\coth \delta_*) (4-\coth \delta_* (\coth \delta_*+4))}\,.
\end{eqnarray}
We can re-introduce now the parameter $\alpha = \frac{\beta \varphi}{4 \pi i}$, and \eqref{FE_exp_mix1} becomes\footnote{\label{footenotecharges} Here, in order to make contact with the field theory quantities, we have followed the same strategy as \cite{BenettiGenolini:2023rkq}, where the bulk potentials $\Phi_e$ are defined to be twice as those in the CFT side $\Phi_e = 2 \Phi$ (notice that in our setup of minimal gauged supergravity we have $\Phi_1=\Phi_2=\Phi_3=\Phi_4=\Phi$). The charge is rescaled accordingly, i.e. $Q_* = 2R_*$, in such a way that $R_*$ and $R_4$ introduced in Sec. \ref{subsec4.1} are quantized in the same way. }
\begin{equation} \label{FE_exp_mix}
    I_{ME} (\beta, \mathfrak{j}, \alpha) =  -S_* - 4 \pi i \alpha R_* - \frac{2\pi^2}{ \beta M_{gap}} \left( 1-4 \alpha^2  \right)\,,
\end{equation}
which is of the desired form. We find that the mass gap parameter $M_{gap}$ is proportional to the quantity $C_T /T$ found in \cite{Larsen:2020lhg} upon setting $g=1$.

\section{The spectrum of quantum AdS$_4$ Kerr-Newman}
\label{sec:quantumKN4}
\subsection{Extracting the density of states}
From the previous section we determined the data to characterize the effective near-horizon theory. Following the procedure in section \ref{subs:sumOversaddles} using \eqref{eq:mixedZ} and \eqref{FE_exp_mix}, we can combine this classical analysis with the results for the quantum corrections coming from the Schwarzian mode, obtaining the following form for the corrected partition function:
\begin{equation} \label{ZJT2}
Z = e^{4 \pi i \alpha R_*} \sum_{n \in \mathbb{Z}} e^{i n \vartheta} \left( \frac{2 \cos(\pi(\alpha + n))}{\pi(1-4(\alpha +n)^2)} \right) e^{S_* + \frac{2\ \pi^2}{\beta M_{gap}} (1-4 (\alpha+n)^2)} \,,
\end{equation}
where we singled out in round brackets the contribution of the 1-loop determinant \cite{Heydeman:2020hhw}. Here, we used the form of the 1-loop determinants first computed in \cite{Stanford:2017thb,Mertens:2017mtv}. In order to treat the general case we have kept the parameter $\vartheta$\footnote{As explained at the end of Sec. \ref{subs:sumOversaddles}, for the rotating supersymmetric black holes $\vartheta$ is set to zero.}.

As in \cite{Boruch:2022tno}, we can find the density of states via the Laplace transform of the quantum corrected partition function. If we work at fixed chemical potential, the states organize into representations of $\mathcal{N}=2$ supersymmetry, so the quantity we first extract is the density of supermultiplets, which is obtained from
\begin{equation} \label{Lapl_trans}
Z (\beta, \mathfrak{j}, \alpha) = \int dE \, e^{-\beta E} \rho(\alpha, \mathfrak{j},E) \,.
\end{equation}
We will evaluate this integral in several steps, consisting in a series of algebraic manipulations.  First of all, via the trigonometric identity
\begin{equation}
    \cos(\pi(\alpha + n)) = (-1)^n \cos(\pi \alpha) = e^{i \pi n} \cos(\pi \alpha) \, , 
\end{equation}
we arrive at
\begin{equation}
Z = 2 e^{4 \pi i \alpha R_*}  \cos(\pi \alpha) \, e^{S_*} \sum_{n \in \mathbb{Z}} \left( \frac{e^{i n (\vartheta + \pi)}}{\pi(1-4(\alpha +n)^2)} \right) e^{ \frac{2\ \pi^2}{\beta M_{gap}} (1-4 (\alpha+n)^2)} \,.
\end{equation}
The integral transform to obtain the density of supermultiplets is simplest if we first differentiate with respect to $\beta$, in order to be able to use some useful identities. We get 
\begin{equation}
\frac{\partial Z} {\partial \beta} = - \frac{4 \pi}{\beta^2 M_{gap}} e^{4 \pi i \alpha R_*}  \cos(\pi \alpha) \, e^{S_*} \sum_{n \in \mathbb{Z}} e^{i n (\vartheta + \pi)}  e^{ \frac{2\ \pi^2}{\beta M_{gap}} (1-4 (\alpha+n)^2)} \,.
\end{equation}
We now Poisson resum 
\begin{equation}
\sum_{n=-\infty}^{\infty} f(n) = \sum_{k=-\infty}^{\infty}\tilde{f}(k) \, , 
\end{equation}
where $\tilde{f}(k) = \int_{-\infty}^{\infty} dn f(n) e^{-2 \pi i k n} $. We obtain 
\begin{equation}
\frac{\partial Z} {\partial \beta} = - \frac{2 \pi}{\beta^2 M_{gap}} e^{4 \pi i \alpha R_*}  \cos(\pi \alpha) \, e^{S_*} e^{ \frac{2\ \pi^2}{\beta M_{gap}} } \sum_{m \in \mathbb{Z} + \frac12 - \frac{\vartheta}{2\pi}} \left( \frac{e^{2 \pi i \alpha m}}{\sqrt{\frac{2 \pi}{M_{gap} \beta}}} \right)  e^{-\frac{\beta m^2 M_{gap}}{8}}\,,
\end{equation}
where we have defined
\begin{equation}
k \equiv m + \frac12 +\frac{\vartheta}{2 \pi} \, , 
\end{equation}
to make the expression more compact. We then use following relation \cite{Stanford:2017thb}
\begin{equation}
\sqrt{\frac{2\pi}{M_{gap}}}  \beta^{-3/2} e^{2\pi^2/ (\beta M_{gap})}= \int_0^{\infty} ds \, e^{-\beta s} \frac{\sinh(2\pi \sqrt{\frac{2s}{M_{gap}}})}{\pi} \, , 
\end{equation}
and we arrive at 
\begin{equation}
\frac{\partial Z} {\partial \beta} = - e^{4 \pi i \alpha R_*}  \cos(\pi \alpha) e^{S_*} \sum_{m \, \in \, \mathbb{Z} + \frac12 - \frac{\vartheta}{2\pi}} e^{2 \pi i \alpha m} \int_0^{\infty} ds \, e^{-\beta \left( s + \frac{m^2 M_{gap}}{8} \right)} \left(\frac{\sinh \left( 2 \pi \sqrt{\frac{2s}{M_{gap}}} \right)}{\pi } \right)\,. \end{equation}
The integral transform has introduced a shift of the energy, so we may redefine 
\begin{equation}
E = s + \frac{m^2 M_{gap}}{8} \qquad dE = ds \, , 
\end{equation}
in order to get
\begin{equation}\label{der_beta}
\frac{\partial Z} {\partial \beta} = - e^{4 \pi i \alpha R_*}  \cos(\pi \alpha) e^{S_*} \sum_{m \, \in \, \mathbb{Z} + \frac12 - \frac{\vartheta}{2\pi}} e^{2 \pi i \alpha m} \int_{\frac{m^2 M_{gap}}{8}}^{\infty} dE e^{-\beta E} \frac{\sinh \left( 2 \pi \sqrt{\frac{2 (E-m^2 M_{gap}/8)}{M_{gap}}} \right)}{\pi } \, ,
\end{equation}
and taking into account $\cos(\pi \alpha) = \frac{e^{i\pi \alpha}+ e^{-i \pi \alpha}}{2}$ we obtain
\begin{equation}
\frac{\partial Z} {\partial \beta} = -   \frac{e^{S_*}}{2} e^{4 \pi i \alpha R_*}  \sum_{m \in \, \mathbb{Z} + \frac12 - \frac{\vartheta}{2\pi}} (e^{ \pi i \alpha (1+2 m)} +e^{ - i \pi \alpha (1-2 m)} ) \int_{\frac{m^2 M_{gap}}{8}}^{\infty} dE e^{-\beta E} \frac{\sinh \left( 2 \pi \sqrt{\frac{2 (E-m^2 M_{gap}/8)}{M_{gap}}} \right)}{\pi }  \, .
\end{equation}

We are ready now to integrate over $\beta$, allowing us to obtain the following expression for the partition function (up to an overall integration constant we determine shortly): 
\begin{equation} \label{final_Z}
Z = Z_{\beta \rightarrow \infty} \delta_{Z_{Sch},0} + e^{S_*} e^{4 \pi i \alpha R_*}  \sum_{Z_{Sch} \in \mathbb{Z} - \frac{\vartheta}{2\pi}} (e^{2\pi i \alpha Z_{Sch}} + e^{2 \pi i \alpha (Z_{Sch}-1)}) \int_{E_{gap}}^{\infty} dE e^{-\beta E} \frac{\sinh \left( 2 \pi \sqrt{\frac{2 (E-E_{gap})}{M_{gap}}} \right)}{2 \pi E}
\end{equation}
from which we can read off the density of multiplets via \eqref{Lapl_trans}. In expression \eqref{final_Z}, we defined $Z_{Sch} = m + \frac12 $ and the quantity
\begin{equation}
E_{gap} = \frac{M_{gap}}{8} \left(Z_{Sch} -\frac12 \right)^2 \,,
\end{equation}
where $E$ and $Z_{Sch}$ are the microcanonical energy and charge of the effective Schwarzian theory arising in the IR. To compute the contribution from BPS states (the first term in \eqref{final_Z}) we send $\beta \rightarrow \infty$ 
 and pass to the microcanonical ensemble, as in \cite{Turiaci:2023jfa}. Starting from \eqref{ZJT2} we have 
\begin{equation}
Z(\beta \rightarrow \infty, \alpha) = e^{4 \pi i \alpha R_*} e^{S_*} \sum_{n \in \mathbb{Z}} e^{i n \vartheta} \frac{2 \cos(\pi(\alpha + n))}{\pi(1-4(\alpha +n)^2)}   = e^{4 \pi i \alpha R_*} e^{S_*} \sum_{n \in \mathbb{Z}}e^{i n \vartheta} \frac{2 (-1)^n \cos(\pi \alpha)}{\pi(1-4(\alpha +n)^2)}  \,.
\end{equation}
We would like to rewrite this formula in a more useful form as a sum over charges. The sum over windings is conjugate to this sum over charges via Poisson resummation, which after a change of variable naively gives
\begin{equation}
\sum_{n \in \mathbb{Z}} e^{i n \vartheta} \frac{2 (-1)^n \cos(\pi \alpha)}{\pi(1-4(\alpha +n)^2)}   =  -\! \! \sum_{k \in \mathbb{Z}-\frac{\vartheta}{2\pi}}\! \! \cos(\pi k)\Theta(2k-1)\left( e^{2\pi i \alpha k}+e^{2\pi i \alpha (k-1)}\right) \,,
\end{equation}
where $\Theta(x)$ is the Heaviside theta-function. This formula requires a few comments. The first is that the sum is taken over an positive integer lattice, possibly shifted by the presence of a $\vartheta$ angle which we argue vanishes for the case of the ABJM. The second is that in the state counting interpretation of this formula, we appear to see contribution of ground states with charges $k$ and $k-1$. These are naively long multiplets related by supersymmetry, in conflict to the claim that we are trying to count short BPS multiplets. However, the sum is actually oscillatory with all but a single term cancelling. Rearranging the sum gives:
\begin{equation}
\sum_{n \in \mathbb{Z}} e^{i n \vartheta} \frac{2 (-1)^n \cos(\pi \alpha)}{\pi(1-4(\alpha +n)^2)}   =  \sum_{k \in \mathbb{Z}-\frac{\vartheta}{2\pi}, \, |k|<\frac12} \! \! \cos(\pi k) e^{2\pi i \alpha k} \,.
\end{equation}
We now have a single set of short multiplets. When $\vartheta = 0$, the only ground state contribution comes from BPS states with $k=0$, and including the exponentially large prefactor gives:
\begin{align}
    Z(\beta \rightarrow \infty, \alpha, \vartheta = 0) = e^{4 \pi i \alpha R_*} e^{S_*} \, .
\end{align}

The final result for the Laplace transform, from \eqref{final_Z}, gives the total density of supermultiplets:
\be \label{density1}
\rho (\alpha, j, E) =e^{4 \pi i \alpha R_*} e^{S_*} \left(   \delta_{Z_{Sch},0} + \sum_{Z_{Sch} \in \mathbb{Z}} (e^{2\pi i \alpha Z_{Sch}} + e^{2 \pi i \alpha (Z_{Sch}-1)})  \frac{\sinh \left( 2 \pi \sqrt{\frac{2 (E-E_{gap})}{M_{gap}}} \right)}{2 \pi E} \Theta(E-E_{gap})\right)\,.
\ee

\subsection{The near-BPS spectrum of ABJM}
We are now in position to make contact with the ABJM field theory dual. We first recall our conventions relating the ADM charge of the black hole, $Q$, with the infrared R-charge $R$ as explained in Footnote~\ref{footenotecharges}. Further, it is more useful to define the scaling dimension $\Delta$ of the extremal BPS states in ABJM as 
\begin{equation} \label{dim_BPS}
\Delta_{BPS} = \Delta (R_*) = \mathfrak{j}+ R_* \,,
\end{equation}
where we have defined
\begin{equation}
\Delta (R) = \mathfrak{j}+  R\,.
\end{equation}
The Schwarzian energy can be related to the scaling dimension $\Delta$ as
\begin{eqnarray} \label{Deltas}
    \Delta &= & \Delta_{BPS} +E_{Sch} +  (R-R_*) = \Delta_{BPS} +E_{Sch} +  Z_{Sch} \,,
\end{eqnarray}
where in the last passage we have used that
\be \label{Rcharge}
R = R_* +Z_{Sch} \,.
\ee

This provides us a way to identify the density of states \eqref{density1} in terms of the spectrum of nearly 1/4 BPS states in ABJM. First of all, the BPS states have $Z_{sch} =0$, hence $R=R_*$ and $\Delta= \Delta_{BPS}$. They correspond to the first term in \eqref{density1}, with degeneracy equal to $e^{S_*}$. The superconformal index corresponds to a value of $\alpha =1/2$, and for this value the second term in \eqref{density1} does not contribute since the sum of the two terms with $Z_{Sch}$ in round bracket  is zero. Notice that the partition function is 
\be
\label{eq:ZJTtoIndex}
Z(\beta, \mathfrak{j}, \alpha = 1/2) = \mathcal{I} (\beta, \mathfrak{j}) = (-1)^{2R_*} e^{S_*} \,.
\ee
 Therefore we have that the BPS states are bosonic if $R_*$ is integer, and fermionic if $R_*$ is half-integer, compatibly with eq. \eqref{spin_stat_CFT}. There are no cancellation in the index, which indeed reproduces the BPS extremal entropy\footnote{Let us mention that logarithmic corrections to the entropy of extremal ($T=0$) AdS$_4$ supersymmetric rotating black holes, for setups of M5-branes wrapped on a hyperbolic 3-manifolds and theories arising on the worldvolume of M2-branes, were investigated respectively in \cite{Benini:2019dyp} and \cite{Bobev:2023dwx}.}. 

 The spectrum of states in ABJM for charge $R= R_*$ come from BPS extremal black hole states with $Z_{Sch} = 0$ (first term in \eqref{density1}) but also from supermultiplets in the second line, which according to \eqref{density1} appear when
\be
E_{Sch} = E_{gap} (Z_{Sch} =0) = \frac{M_{gap}}{32} \,,
\ee
hence, translating in conformal dimensions via \eqref{Deltas}
\be
\Delta > \Delta_{BPS}+ \Delta_{gap} \, , \qquad \text{with} \qquad \Delta_{gap} \equiv \frac{M_{gap}}{32} \,.
\ee
In other words, $\Delta_{gap}$ is \textit{the gap between the 1/4 BPS extremal black hole and the lightest excited state with charge $R_*$}. Using the result from \eqref{MGAP}, we can re-express the gap in terms of field theory quantities, using the relation $\frac{1}{G} = \frac{2 \sqrt2}{3} N^{3/2}$ for ABJM, as
\begin{eqnarray}
\Delta_{gap}  & = & \frac{G (2-\coth \delta_*) (4-\coth \delta_* (\coth \delta_*+4))}{32(\coth \delta_* -1)^{3/2}} \nonumber \\
& = &\frac{3}{64 \sqrt2 N^{3/2}} \frac{ (2-\coth \delta_*) (4-\coth \delta_* (\coth \delta_*+4))}{(\coth \delta_* -1)^{3/2}} \,.
\end{eqnarray}
The density of states of charge $R_*$, written again in field theory language, is obtained by Fourier transforming \eqref{density1} to this charge sector:
\be
\label{eq:BPSDOS}
\rho (\Delta, \mathfrak{j}, R_*) = e^{S_*} \delta (\Delta -\Delta_{BPS}) + \frac{e^{S_*} \sinh \left( \pi \sqrt{\frac{\Delta -\Delta_{BPS} -\Delta_{gap}}{4 \Delta_{gap} }} \right)}{\pi (\Delta -\Delta_{BPS})} \Theta (\Delta - \Delta_{BPS} -\Delta_{gap})\,.
\ee
The density of states, for a sample value of parameters, is displayed in Fig.~\ref{Fig0}(a).

States with $R \neq R_*$ are non-supersymmetric and there is a similar continuous density of states without the delta function contribution. The extremal black holes in this sector thus begin at their respective gap scale. For an extremal black hole, for a certain fixed $R$, states with Schwarzian charge $Z_{sch}$ begin at an energy,
\be
E_{Sch} = E_{gap}(Z_{Sch}) =  \frac{M_{gap}}{32} \left( Z_{Sch} -\frac12 \right)^2 = \frac{M_{gap}}{32} \left( R-R_* -\frac12 \right)^2 \,,
\ee
where we have used \eqref{Rcharge}. At the same time that from $Z_{Sch} +1$
\be
E_{Sch} = E_{gap}(Z_{Sch}) =  \frac{M_{gap}}{32} \left( Z_{Sch} +\frac12 \right)^2 = \frac{M_{gap}}{32} \left( R-R_* +1 \right)^2 \,.
\ee
Thereby, inserting into \eqref{Deltas}, the value of the conformal dimension is found as the minimum
\be \label{DeltanonBPS}
\Delta_{extremal} (R \neq R_*) = min_{\pm} \left[ \Delta_{BPS} + \frac{M_{gap}}{32} \left( R-R_* \pm \frac12 \right)^2  + \frac12 \left( R- R_* + \frac12 \pm \frac12\right)\right] \,,
\ee
where the minimum is for a given value of $R$. Notice that for $R = R^*$ we get only one value of $\Delta$, namely $\Delta_{extremal}(R = R^*) = \Delta_{BPS}$, as expected. The degeneracy for the non-supersymmetric black holes with $R \neq R^*$ at $T=0$ goes to zero, as we will see below. The density of states for $R \neq R_*$ is again a Fourier transform of \eqref{density1}, which is now a sum of two terms from states in different multiplets:
\begin{eqnarray}
\label{eq:nonBPSDOS}
\rho (\Delta, \mathfrak{j}, R) & = & \frac{e^{S_*} \sinh \left( \pi \sqrt{\frac{\Delta -\Delta_{extremal}(R)}{4 \Delta_{gap} }} \right)}{\pi (\Delta -\Delta_{SUSY}(R))} \Theta (\Delta - \Delta_{extremal}(R)) \nonumber \\
 & + & \frac{e^{S_*} \sinh \left( \pi \sqrt{\frac{\Delta -\Delta_{extremal}(R+1)}{4 \Delta_{gap} }} \right)}{\pi (\Delta -\Delta_{SUSY}(R+1))} \Theta (\Delta - \Delta_{extremal}(R+1)) \,,
\end{eqnarray}
where we notice the absence of the Dirac delta function. Plotting it for particular values of parameters yields Fig.~\ref{Fig0}(b), which is qualitatively similar to the density of states for the Kerr solution. One can see the kink in the density of states that signals the onset of the multiplet with $R+1$.

\section{Discussion}

The low temperature corrections taken into account in this work are a unique fingerprint of quantum gravity, and represent an important physical effect that can not be ignored if one wants to interpret a black hole (near extremality) as an ordinary quantum system. In the context of AdS$_4$ black holes with 3d SCFT duals, one can make very precise the quantum system, at least when supersymmetry is preserved \cite{BenettiGenolini:2023rkq}. Our work represents an important first step in extending this correspondence to non-supersymmetric observables for \emph{quantum} black holes in which one must sum over a family of solutions weighted by their fluctuation determinants. Our results for the spectrum are succinctly summarized in \eqref{eq:BPSDOS} when the background electric charge is set to the BPS value, and \eqref{eq:nonBPSDOS} for more general values of the background charge. Supersymmetric black holes and the mass gap are present in the former spectrum, while the latter is characteristic of more generic AdS$_4$ Kerr-Newman black holes without supersymmetry.

We expect our calculations to be valid for sufficiently large $N$ but temperatures polynomially small in $N$; here the fluctuations of the near-horizon region dominate, but exponentially small non-perturbative M-theory/quantum gravity effects are still suppressed. It may be possible to incorporate these non-perturbative corrections in a systematic expansion, similar to the one in \cite{Aharony:2021zkr} for branes, and \cite{Saad:2019lba} for spacetime wormholes.

At leading order in the entropy and temperature, the AdS$_4$ Kerr-Newman black hole and the resulting $\mathcal{N}=2$ Schwarzian mode we considered leads to universal predictions for the spectrum of 3d $\mathcal{N}=2$ SCFT's. This is because the $\mathcal{N}=2$ Schwarzian theory represents a nontrivial EFT, and the parameters of this EFT a determined by embedding into the UV gravity theory. In fact, an important part of this work was to derive which version of the Schwarzian theory is applicable in the case of ABJM. Using the known properties of the UV AdS$_4 \times S^7$ / ABJM duality, our EFT results show that the density of states in most charge sectors go to zero at low energies. In the special charge sector ($R = R^*$ in our notation), our results are consistent with the large $N$ superconformal index of ABJM (including the alternating signs in $R^*$ due to bose/fermi weighting of the index). Beyond this, we confirm the extremal degeneracies are consistent with the absolute value of the index, providing to our knowledge the first proof of this for AdS$_4$ black holes.

Similar to other theories in which the $\mathcal{N} \geq 2$ Schwarzian theories are a useful IR description, we predict a mass gap in operator dimensions for ABJM which grows as $N^{-\frac32}$. While we do not expect this gap to be protected from corrections due to finite gauge couplings \cite{Boruch:2022tno}, it is robust at large $N$ because it is parametrically larger than the microscopic level spacings of non-BPS operators (which are expected to be spaced at order $\exp(-N^\frac32)$). It is possible that a large $N$ path integral computation in the ABJM theory could reproduce this result\footnote{See \cite{Cabo-Bizet:2024gny} for a similar computation in $\mathcal{N}=4$ SYM.}. Alternatively, it may be possible to study the supercharge spectrum for small $N$ as in \cite{Chang:2023zqk}. Both of these suggestions are complicated by the fact that the ABJM partition function includes contributions from non-BPS monopole operators which would lead to complicated anomalous dimensions for multi-particle states. In any event, we feel the gravity computation shows this gap is a sharp quantum effect separating the ground states from the continuum, and thus one expects the BPS sector of ABJM to be nontrivial, even for arbitrarily low energies / late times \cite{Lin:2022zxd,Chen:2024oqv}. The most dramatic possibility along these lines is that ABJM in the deconfined phase becomes well approximated by the supersymmetric random matrix ensemble constructed in \cite{Turiaci:2023jfa}.

We finally comment that these rotating black holes we study are potentially part of a larger phase diagram for AdS$_4$ black holes that contain radiant instabilities\cite{Kim:2023sig,Choi:2024xnv}. These possibilities are invisible in our approach, but may have to be included if one considers charges too far from extremality. In this case, the bulk phase might be better understood as a black hole with scalar hair, or additional dual giant M2 branes. Understanding the full phase diagram and the manifestation in the dual 3d CFT is an interesting problem for future work.

\section*{Acknowledgements}

We thank J. Boruch, A. Castro, K. Hristov, L. Iliesiu, J. Maldacena, S. Murthy, L. Pando Zayas, M. Rangamani, A. Strominger, G. Turiaci, for helpful comments, interesting discussions, and collaborations on related topics. MTH is supported by Harvard University and the Black Hole Initiative, funded in part by the Gordon and Betty Moore Foundation (Grant 8273.01) and the John Templeton Foundation (Grant 62286). MTH completed part of this work while at the Kavli Institute for Theoretical Physics (KITP), supported in part by grant NSF PHY-2309135. The work of CT is supported by the Marie Sklodowska-Curie Global Fellowship (H2020 Program) SPINBHMICRO-101024314. This work was performed in part at Aspen Center for Physics, which is supported by National Science Foundation grant PHY-2210452. CT also acknowledges support from the Simons Center for
Geometry and Physics, Stony Brook University at which some of the research for this paper was performed.

\bibliographystyle{./JHEP-2}
\bibliography{main.bib}

\end{document}